\documentclass[12pt,letterpaper]{article}

\usepackage{scicite}
\usepackage[varg]{txfonts}
\usepackage{graphicx}

\newenvironment{sciabstract}{%
\begin{quote} \bf}
{\end{quote}}

\newcounter{lastnote}
\newenvironment{scilastnote}{%
\setcounter{lastnote}{\value{enumiv}}%
\addtocounter{lastnote}{+1}%
\begin{list}%
{\arabic{lastnote}.}
{\setlength{\leftmargin}{.28in}}
{\setlength{\labelsep}{.5em}}}
{\end{list}}

\title{Lighting the Universe with filaments}

\author
{Liang Gao$^{1, \ast}$, Tom Theuns$^{1,2}$
\\
\\
\normalsize{$1$ Institute for Computational Cosmology, Durham
  University, South Road, Durham DH1 3LE, UK}\\
\normalsize{$2$ Department of Physics, University of Antwerp,
  Groenenborgerlaan 171, B-2020 Antwerpen, Belgium} \\
\normalsize{$^\ast$ E-mail: liang.gao@durham.ac.uk} \\
}
\date{}

\def\gsim{\mathrel{\rlap{\lower 4pt \hbox{\hskip 1pt $\sim$}}\raise 1pt \hbox {$>$}}}
\def\lsim{\mathrel{\rlap{\lower 4pt \hbox{\hskip 1pt $\sim$}}\raise 1pt \hbox {$<$}}}

\begin{document}




\maketitle


\begin{sciabstract}

The first stars in the Universe form when chemically pristine gas heats
as it falls into dark matter potential wells, cools radiatively due to
the formation of molecular hydrogen, and becomes self-gravitating. We
demonstrate with super-computer simulations that their properties
depend critically on the currently unknown nature of the dark matter.
If the dark matter particles have intrinsic velocities that wipe-out
small-scale structure, then the first stars form in filaments with
lengths of order the free-streaming scale, which can be $\sim
10^{20}$~m ($\sim 3$~kpc, baryonic masses $\sim 10^7~$ solar masses)
for realistic \lq warm dark matter\rq~ candidates. Fragmentation of the
filaments forms stars with a range of masses which may explain the
observed peculiar element abundance pattern of extremely metal-poor
stars, while coalescence of fragments and stars during the filament's
ultimate collapse may seed the super massive black holes that lurk in
the centres of most massive galaxies.

\end{sciabstract}

Most of the matter in the Universe does not interact with light except
gravitationally. This \lq dark matter\rq~ is usually assumed to be \lq
cold\rq, meaning that its velocity dispersion is sufficiently small for
density perturbations imprinted in the early Universe to persist up to
very small scales. Although this model is able to describe the
large-scale distribution of galaxies in impressive detail, it may face
problems on the scale of galaxies and below, for example it may predict
too many satellite galaxies \cite{M99}, as well as too cuspy profiles
for the dark matter halos that surround galaxies\cite{DH01}.

However, dark matter has yet to be detected in the laboratory, and
there exist many viable dark matter candidates from particle physics
that are not \lq cold\rq~. Warm dark matter (WDM) particles have
intrinsic thermal velocities and these motions quench the growth of
structure below a \lq free-streaming\rq~ scale (the distance over which
a typical WDM particle travels) which depends on the nature of the
particle. Because small and dense halos do not form below the
free-streaming-scale, the dark matter halos that surround galaxies in a
WDM model have far less substructure and are less concentrated as
compared to their cold dark matter (CDM) counterparts, which may help
alleviate both the satellite and galactic core problem
\cite{Bode01}. Structures on larger scales are similar in WDM and CDM
and therefore the distribution of galaxies is not affected. The first
generation of stars in the Universe forms when primordial gas gets
compressed by falling into small dark matter potential wells
\cite{ABN02, BCL02, Yoshida06, G07}. Because WDM affects structure
formation on such small scales it may influence how the first stars
form; we have performed simulations to analyse this in more detail.

Large-scale power in the spectrum of density perturbations causes
progenitors of present-day clusters of galaxies to be amongst the first
objects to condense out of the initially almost smooth mass
distribution. We study the early formation stages of such an object by
identifying a massive cluster of galaxies in a dark matter simulation
of a large cosmological volume at redshift $z=0$, and use a multi-scale
technique\cite{G05, SO} to re-simulate its formation and evolution with
the cosmological hydrodynamics code Gadget-2 \cite{S05, SO}. Baryons
compressed by falling into the developing dark matter potential wells
cool radiatively through molecular hydrogen emission lines\cite{GP98,
SO}; we follow the formation of molecular hydrogen and its cooling
using a sophisticated chemistry network\cite{Yoshida06}. The parent
simulation, from which we picked the re-simulated region, followed the
growth of structure in a CDM dominated Universe with a cosmological
constant $\Lambda$ ($\Lambda$CDM); we performed the re-simulations
assuming both CDM and WDM as dark matter. In this WDM model, small
scale power in the density fluctuations is exponentially suppressed
below $\sim 100$~co-moving kpc, mimicking free-streaming of gravitino
particles with mass $m_{\rm WDM}=3$~keV\cite{Bode01}. Gravitinos are a
popular WDM candidate\cite{bhs05} but our results are insensitive to
the exact choice of WDM particle as long as its free-streaming length
is more than a few tens of co-moving kilo parsecs. Therefore even if
the gravitino were slightly more massive, or if the WDM were instead a
sterile neutrino (another popular WDM candidate\cite{neutrinos}), our
results would not change appreciably. Observations of the clustering of
neutral gas along sight lines to distant quasars (the Lyman-$\alpha$
forest of absorption lines) probe scales $1-40$~Mpc when the density
perturbations on these scales are still small, and the presence or
absence of significant substructure in these observations can constrain
the masses of WDM particles\cite{Viel06, Seljak06}; our choice of WDM
particle mass (3 keV, the corresponding mass-scale is $\sim 3\times
10^8M_\odot$) is well above this lower limit ($\sim 2$ keV;
\cite{Viel06, Seljak06}). The initial amplitude of the imposed density
perturbations in our simulations is normalised to the level seen in the
Cosmic Microwave Background radiation\cite{wmap3}, and our simulations
start at redshift $z\sim 200$. (See the supplementary online material
for more details on these simulations.)

The growth of structure in these re-simulations leads to a pattern of
filaments and sheets (Fig.1) familiar from the local large-scale
distribution of galaxies. This is because the assumed Gaussian spectrum
of density perturbations, appropriate for an inflationary model, leads
to collapse along one (sheet) and two directions (filament) before the
formation of halos. Although the large-scale filamentary pattern is
very similar in CDM and WDM, the structure of the filaments themselves
is very different: whereas the CDM filaments fragment into numerous
nearly spherical high density regions (\lq halos\rq; panel 1.a), the
WDM filaments are mostly devoid of such substructure (panel
1.b). Panels 1.c and 1.d depict the WDM filament at an earlier time
before any of the other filaments have formed: the central density is
very high (hydrogen number density $n_{\rm h}\sim
10^4\,$cm$^{-3}\approx 10^6\langle n_{\rm h}\rangle$) yet no dark
matter halo has formed yet. It is well known that the Poisson noise in
simulation codes that use particles to represent the dark matter leads
to spurious fragmentation of the filaments that form in such WDM
simulations\cite{Goetz02, W07}. We therefore end the analysis of our
WDM simulations well before filaments fragment.

The length of the filament ($\sim 3$~kpc) is of order of the imposed
WDM free-streaming scale as expected, $L \sim 4$~kpc at redshift
$z=23.34$ when the Universe is $140$ million years old. Gas and dark
matter accrete perpendicular onto the filament's axis (Fig.~2). Dark
matter particles falling into the filament perform damped oscillations
as the potential well deepens. At $r\sim 50$~pc (where $r$ is the
distance perpendicular to the filament's axis) dark matter particles
falling into the well encounter particles that fell in from the other
side, and such successive instances of \lq orbit-crossing\rq~ give rise
to the steps in the density seen in Fig.~1b.  Baryons do not undergo
orbit-crossing but the gas gets compressed to a temperature $T\approx
7000$ K at $r\sim 20$~pc. Rapid build-up of ${\rm H}_2$ induces cooling
and the gas starts to dominate the matter density further downstream,
so that the ratio of gas to dark matter densities $\rho_b/\rho_{\rm
DM}\sim 15=100\langle\rho_{\rm b}\rangle/\langle\rho_{\rm DM}\rangle$
at $r=2$~pc. At $r<2$ pc where the gas dominates, the ratios of
principal axes of the filament are $b/a=0.123$ and $c/a=0.118$ hence
the filament is very nearly cylindrical. The properties of both the gas
and the dark matter are very uniform along the whole length of the
filament. The cylindrical density profile below $10$~pc is
approximately $\rho \propto r^{-2.8}$ for $2\le r\le 8$~pc and $\rho
\propto r^{-2}$ for $r<2$~pc. This contrasts with $\rho \propto
r^{-2.3}$ for the spherically averaged profile of the gas in CDM halos
on a comparable scale \cite{Yoshida06, G07}. The central ${\rm H}_2$
abundance reaches $10^{-3}$, higher than in the CDM case because of the
higher temperature reached behind the accretion shock. This higher
temperature, and the associated higher ionisation fraction, will also
enhance the importance of HD cooling at later stages \cite{HD,
Yoshida07}.

The non-linear collapse into a thin filament found in these WDM
simulations is in sharp contrast to what happens in the CDM case.
There the first objects to reach high densities are discrete, nearly
spherical, dark matter halos (gravitationally bound concentrations of
dark matter) that form at tiny masses and build-up hierarchically
through mergers and accretion. Some halos have a sufficiently deep
potential well to accrete and shock baryons, enabling ${\rm H}_2$
formation and radiative cooling. Runaway collapse of the rapidly
accreting self-gravitating gas is thought to lead to the formation of a
single massive star per cooling halo \cite{ABN02, Yoshida06, G07}. The
absence of small scale power in WDM prevents halos from forming before
the filament itself becomes highly non-linear.  Although our
re-simulations focus on the progenitor of a massive cluster, and hence
the forming object collapses unusually early on, the fact that very
high densities are reached in a filament, as opposed to a spherical
halo, is generic to WDM.

The stability of collapsing filamentary clouds has been investigated in
the context of the formation of cloud cores by for example\cite{L85,
IM97}, and applied to early Universe filaments \cite{NU02}. The
inability of gas to cool sufficiently fast usually limits the collapse
time of the filament by the cooling time, $t_d\equiv\rho/\dot\rho\sim
t_c\equiv nk_{\rm Boltz}T/\Lambda$ (where $T$ is the temperature and
$\Lambda$ is the cooling rate due to H$_2$ and HD
cooling). Perturbations start growing, possibly leading to
fragmentation, when the dynamical time $t_d\gg t_p$, where $t_p$ is the
inverse growth rate of perturbations. If gas cooling is efficient,
$t_d\sim t_c\ll t_p$, and perturbations do not grow. At densities
$n=n_1\sim 10^5$~cm$^{-3}$ the level population of H$_2$ reaches LTE
making $\Lambda\propto n$ instead of $\propto n^2$ and $t_d$
increases\cite{GP98}. A sufficiently massive filament may yet survive
fragmentation at this stage, and at higher density $n\sim
10^9$~cm$^{-3}$ three-body processes promote the formation of H$_2$ and
the cooling time decreases again. However when $n\geq n_2=
10^{12}$~cm$^{-3}$ the gas becomes optically thick in the H$_2$ cooling
transitions slowing the collapse and the filament is once more in
danger of fragmenting. Collision induced continuum emission will again
decrease the cooling time when $n\sim 10^{14}$~cm$^{-3}$, until the gas
becomes optically thick also to this cooling radiation at densities
$n\geq n_3= 10^{16}$~cm$^{-3}$ \cite{Omukai98, Yoshida07}. The physics
of the H$_2$ molecule therefore sets three densities at which the
filament may fragment. The typical fragment masses are of order tens of
solar masses, solar masses, and sub-solar masses, for fragmentation at
densities $n_1$, $n_2$ and $n_3$, respectively\cite{NU02}.

The tidal field around the filament breaks the cylindrical symmetry on
scales comparable to the filament's length (Fig.~1), and can trigger
the gas dynamical instabilities that ultimately lead to
fragmentation. Unfortunately our current simulations are not able to
follow this process in detail because these tiny deviations from
symmetry are overpowered by numerical noise caused by the graininess of
the particle distribution. This causes the filament to fragment very
rapidly as expected, yet the scale of the fragmentation is
artificial. In the WDM universe, the small-scale perturbations that
trigger fragmentation in the simulations are not present and need to be
generated through transfer of power from larger scales. Since this is a
relatively slow process, the central density will have reached the
higher value $n_2$ or even higher, implying small fragment masses of
order of a solar mass or below. Such fragments can coalesce to form
more massive clumps, as demonstrated in 2D numerical
simulations\cite{IM97}. Even if individual cores survive, they may
still grow in mass through accretion.

Detailed observations of star-forming clouds in the Milky Way reveal
that the low-mass stellar mass function is very similar to that of the
dense, pre-stellar cloud cores within the ambient cloud, although it is
not yet clear what determines the cut-off at high masses. This close
similarity might indicate that the process of cloud fragmentation plays
an important role in determining the initial mass
function\cite{Larson05}. If this also applies to star-formation in a
WDM filament, then it is plausible that fragmentation will lead to a
burst of star formation which includes low-mass stars with masses $\sim
1~M_\odot$ or below, but also much more massive stars built through
mergers and accretion. The more massive stars will end their short
lives through supernova explosions or collapse to form intermediate
mass black holes, and some of the very low mass stars may potentially
survive until today. Although the details of this scenario are
uncertain, it is clear that the stellar mass function will be quite
different from the CDM case.

The low-mass Milky Way stars HE 0107-5240 \cite{CH02} and HE 1327-2326
\cite{F05} have extremely low metallicities\footnote{The notation [X]
$\equiv \log({\rm X}/{\rm X}_0)$, characterises the mass fraction of
element X in terms of the value in the Sun, X$_0$.}  of [Fe/H] $\sim$
$-5$ and peculiar element abundances, for example [C/Fe]$>$1. An
initial mass function of zero metallicity first stars with a range of
masses, as we suggested might be the case in WDM, could explain these
stars either as due to self-pollution or pre-enrichment by intermediate
mass (tens of solar masses) supernovae\cite{BC05}. Such first
generation low and intermediate mass stars are not thought to form in
CDM\cite{ABN02, Yoshida06, G07}.  Therefore, although the present
observational evidence is not yet unambiguous, a future detection of
zero-metallicity low-mass first stars may indicate that the dark matter
is warm.

Free-streaming of the WDM decreases significantly the number density
of halos that host early star formation, which could delay
reionization as compared to CDM\cite{Y03}. However the additional
mode of star formation in filaments could partly compensate, making
it unclear whether reionization is indeed delayed when the dark
matter is warm.

What is the ultimate fate of such $\sim 10^7\,M_\odot$ filaments?
Eventually the filament will collapse along its long axis, and since
its mean density is very high a significant number of collisions
between cloud cores and stars would appear inevitable. Such collisions
could build-up a massive object which can seed the formation of the
super-massive black holes that power redshift $z \sim 6$ quasars, and
appear to lurk in the centers of most large galaxies today.

The different outcome of WDM versus CDM is because the thermal
velocities of the WDM particles prevent halos from forming before the
filaments themselves form stars. This happens when the free-streaming
length introduced by the dark matter's thermal velocities is more than
a few tens of co-moving kilo parsecs so that the filaments are
sufficiently massive to heat infalling gas to more than $\sim $1000~K,
enabling efficient cooling by molecular hydrogen. If we simulate a WDM
Universe with much shorter free-streaming length, for example $\sim
20$~kpc corresponding to a gravitino mass $m_{\rm WDM}=15$~keV, star
formation proceeds in a similar way to the CDM case, in good agreement
with earlier work\cite{ON06}. The likely very different initial mass
function of the first stars and the rapid formation of massive black
holes in a WDM scenario with $m_{\rm WDM}\sim 3$~keV, as opposed to
CDM, implies a very different early thermal and metal enrichment
history of the Universe, greatly affecting subsequent galaxy
formation. It appears therefore that the way in which quasar, star and
galaxy formation started, depends strongly on the nature of the dark
matter.


\begin{scilastnote}
\item It is a pleasure to thank Carlos Frenk, Richard Bower, Cedric Lacey,
Adrian Jenkins, Shude Mao, Volker Springel and Naoki Yoshida for
helpful discussions. All computations were performed on the Cosmology
machine of the Institute for Computational Cosmology at Durham
University. This work was supported by a PPARC rolling grant.
\end{scilastnote}

\clearpage

\begin{figure*}
\resizebox{14cm}{!}{\includegraphics{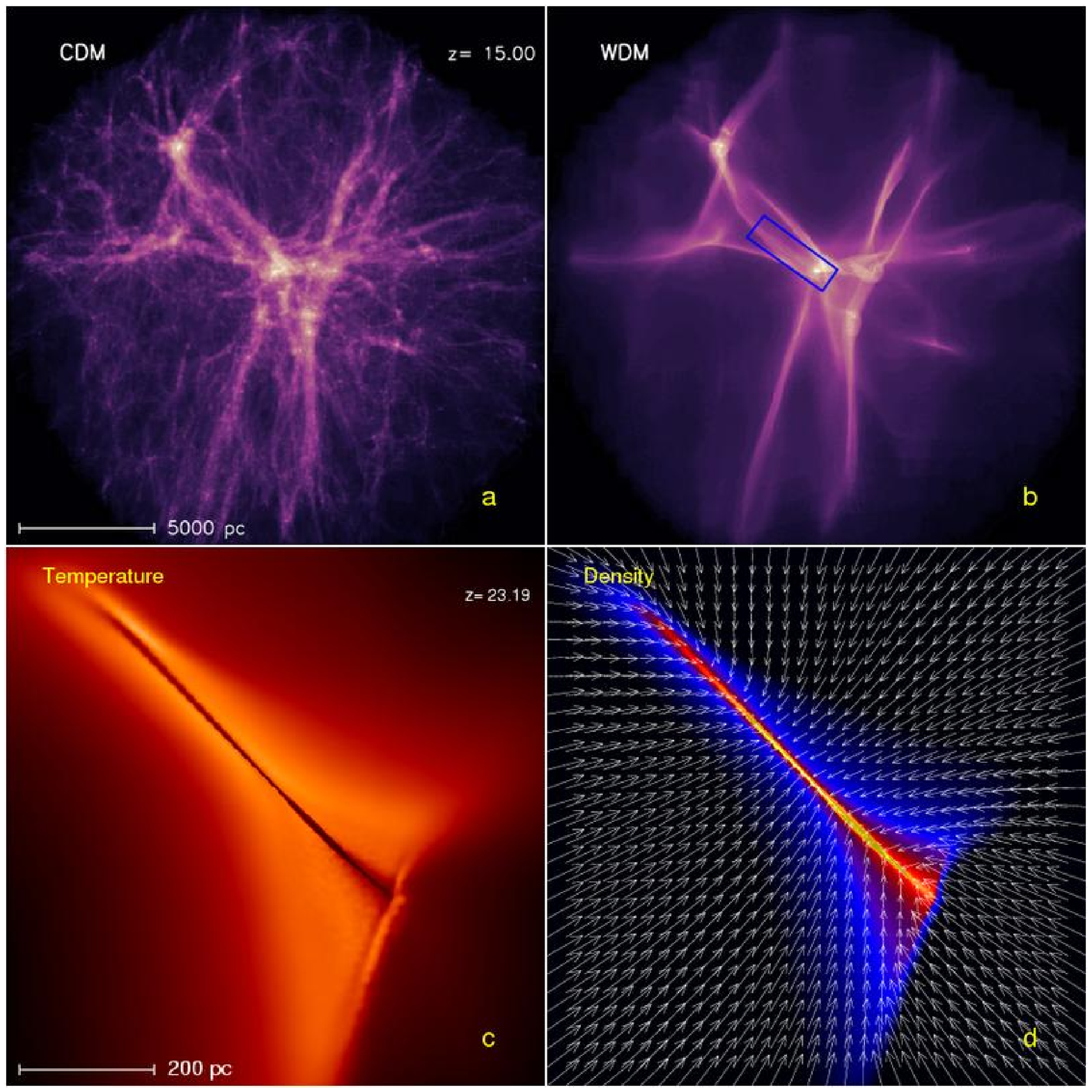}}
\label{fig:filaments}
\end{figure*}
\noindent Figure 1\\ {\it Top panels:} Dark matter density structure of
the progenitor of a redshift $z=0$ massive cluster of galaxies at
$z=15$ when the thermal velocities of the dark matter particles are
negligible (CDM; panel a) and when the dark matter is \lq warm\rq~
(WDM; a gravitino with $m_{\rm WDM}=3$~keV; panel b). Although both
models produce a characteristic filamentary pattern in the density, the
CDM filaments fragment into numerous nearly spherical halos whereas
free-streaming of the WDM prevents such substructure from forming.
{\it Bottom panels:} Gas temperature (panel c) and density (panel d) in
the WDM filament indicated by the red box in panel b, at an earlier
redshift ($z=23.34$) when only this filament had formed.  Gas accretes
very uniformly onto the filament as indicated by the velocity vectors,
heats as it gets compressed, but further downstream cools due to the
formation of H$_2$, making the centre of the filament cold and
dense. The filament shown here is almost perfectly cylindrical, more
generally they are elliptical in shape.
\clearpage

\begin{figure*}
\hspace{0.13cm}\resizebox{8cm}{!}{\includegraphics{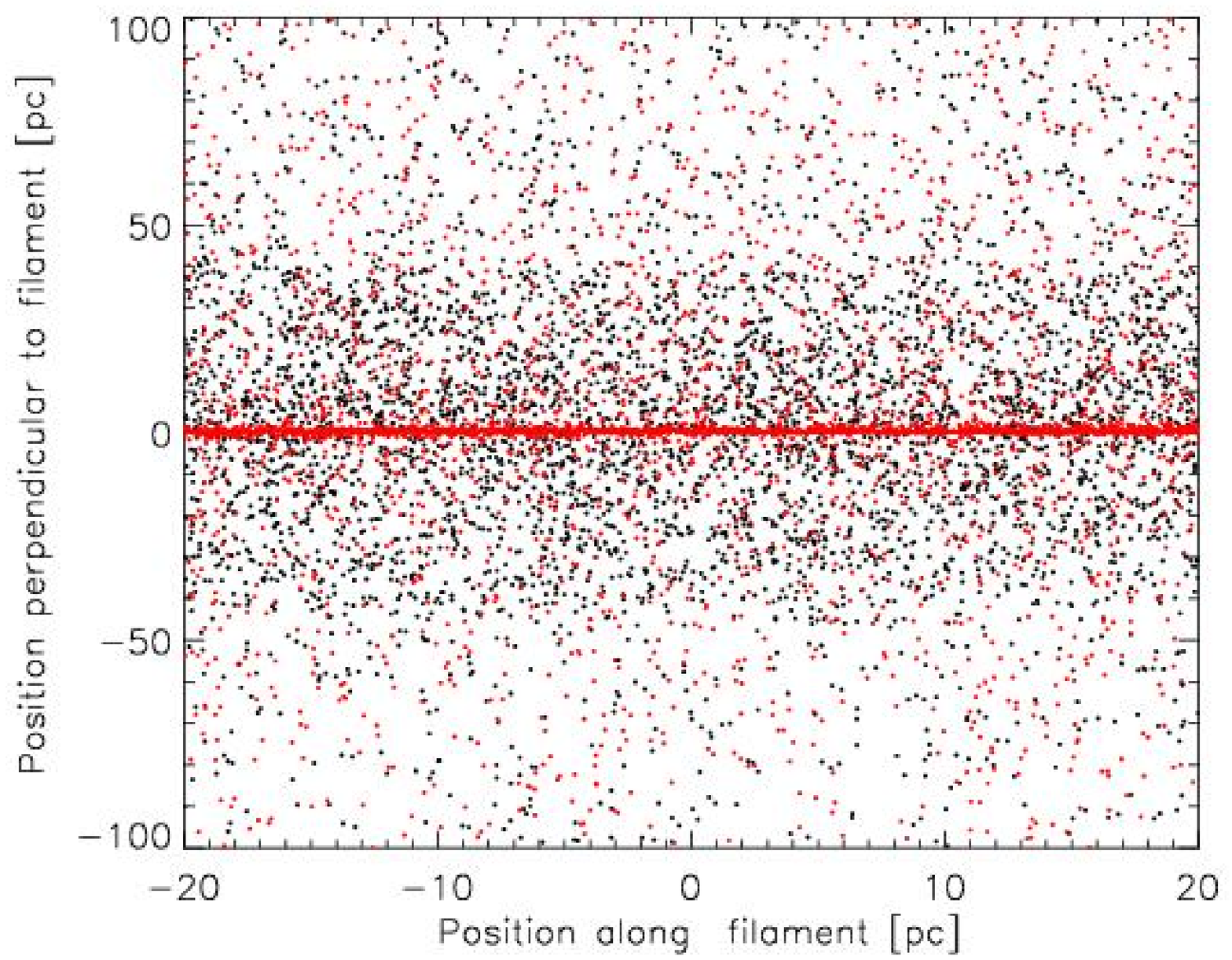}}
\hspace{0.13cm}\resizebox{8cm}{!}{\includegraphics{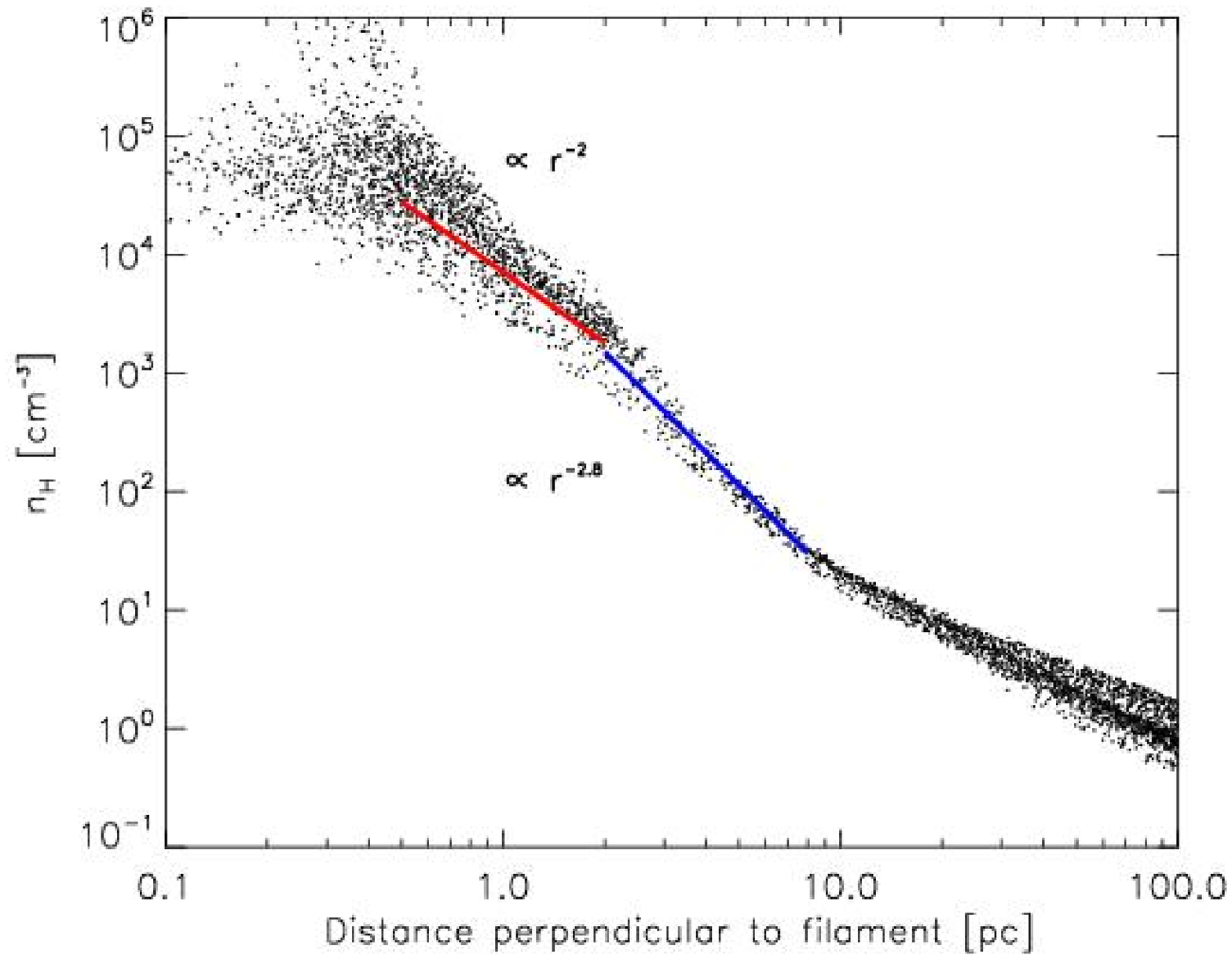}}
\hspace{0.13cm}\resizebox{8cm}{!}{\includegraphics{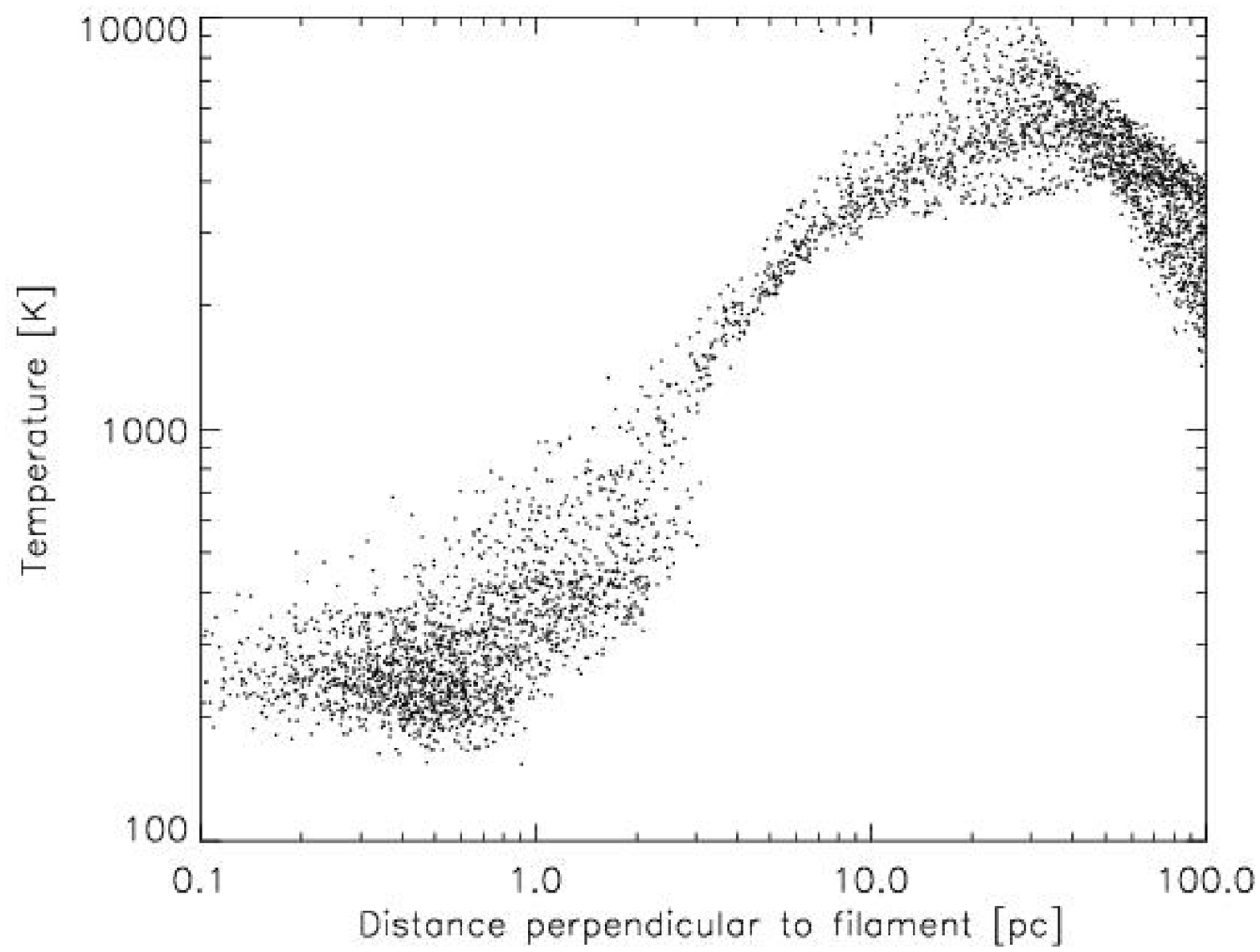}}
\hspace{0.13cm}\resizebox{8cm}{!}{\includegraphics{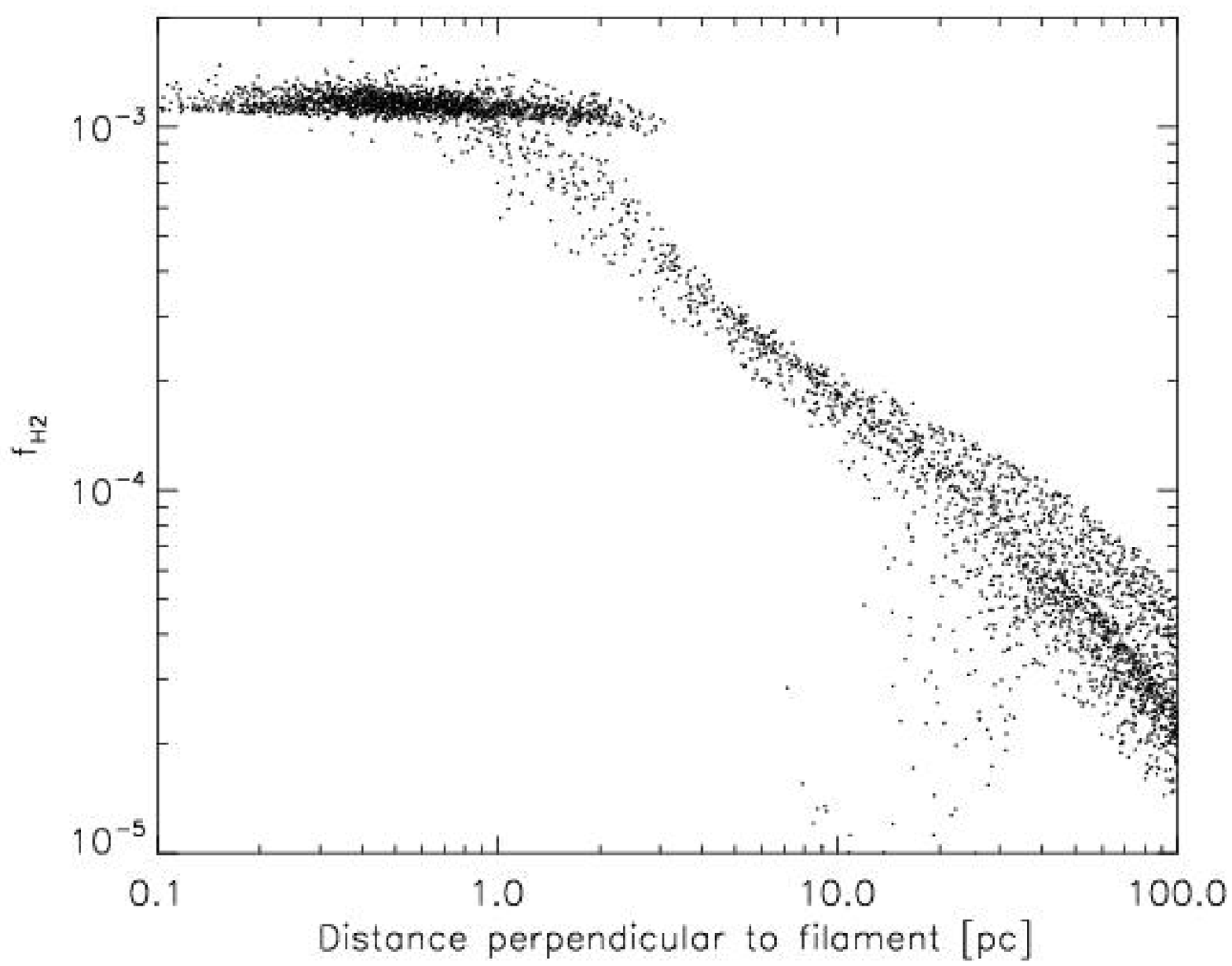}}
\label{fig:fig2}
\end{figure*}
\noindent Figure 2\\ Gas and dark matter profiles of the WDM filament
shown in panels c and d of Fig.~1. Top left: accretion of dark matter
particles (black) and gas particles (red); note the different
horizontal and vertical scales (along and perpendicular to the
filament, respectively). The dark matter undergoes orbit-crossing
(multi-valued velocities) at $r\sim 50$~pc, whereas the gas gets
shocked much closer in ($\sim 20$~pc) and dominates further
downstream. Top right: gas density profile as measured perpendicular to
the filament's axis. Lines $\rho\propto r^{-2}$ and $\rho\propto
r^{-2.8}$ are drawn to guide the eye. Bottom left: temperature profile:
gas heats as it gets compressed, but the rapid build-up of ${\rm H}_2$
cools the gas further in. Bottom right: Build-up of molecular hydrogen
fraction, $f_{H_2}$, during the accretion.


\newpage
{\bf Supporting Online Material}

Material and Methods

SOM text

Figs:S1, S2, S3

\newpage


\begin{center}
\vspace*{-10.00pt}
{\Large \bf SUPPORTING ONLINE MATERIAL}
\end{center}


\noindent We briefly describe the setup of the initial conditions, how
we model the effect of warm dark matter, the chemistry network, and the
numerical method used.

\section{Numerical Methods}
\subsection{Cosmological initial conditions}
The theory of inflation posits that quantum mechanical fluctuations in
the \lq inflaton\rq\ field were blown-up to macroscopic scales during a
short period of rapid expansion called inflation. These small
perturbations in the density have grown due to gravity into the
large-scale structure we see today ${\it(S1)}$. The initial conditions
of a cosmological simulation represent these initially small
perturbations at the starting redshift of the simulation in a periodic,
computational box of given co-moving size. The simulation then follows
how these small ripples develop into high and low-density regions.

If the statistics of the perturbations are those of a Gaussian field,
as is thought to be the case in inflation, then the initial
fluctuations are fully described by their power-spectrum, $P(k,t)$,
which characterises the amplitude of the fluctuations as function of
scale, $\lambda=2\pi/k$, at time $t$. The Warm Dark Matter (WDM) and
Cold Dark Matter (CDM) power spectra from {\it (S2)} are plotted in
Fig.~S1: free-streaming of a WDM particle with mass $m_{\rm WDM}=3$~keV
suppresses power below a scale $\sim 100$~kpc as compared to CDM.

Setting-up initial conditions now boils down to generating a Gaussian
field with known power-spectrum, which is done using Fourier transform
methods. Since our simulation code uses particles to represent the
matter distribution, we need to generate a particle distribution to
represent this density field. We start by creating a homogeneous
particle distribution, either by simply putting equal mass particles on
the vertices of a regular grid, or as done here, starting from a glass
distribution. We then use the Zel'dovich approximation ${\it(S3)}$ to
displace these particles to represent the density field (see also ${\it
S4}$). The physical power-spectrum has power on all scales, but the
particle distribution cannot represent waves larger than the box size,
or smaller than twice the inter particle spacing $d$.

The dark matter structures that host the very first stars are very rare
objects, therefore cosmological simulations of such early structure
formation must model extremely large volumes in order to sample these
rare objects. However, the simulations must also achieve very high mass
resolution to follow accurately how gas collapses to high densities in
these early potential wells. The contradictory requirements of large
simulation volume yet high resolution make cosmological simulation of
these structures unfeasible with uniform mass resolution given current
computational resources.  Therefore we use a \lq multi-scale
re-simulation technique\rq\ to overcome this dilemma. In a
re-simulation, only a sub-volume of the whole simulation volume
(containing the object in which we are interested) is simulated with
very high resolution, while the rest of the computational domain is
sampled at much coarser resolution. The advantage is that it enables us
to obtain the extremely high dynamic range in mass and length required
for the current problem; the challenge is how to generate the
appropriate initial conditions (ICs).

We begin by generating ICs in a large periodic box (479 Mpc, the \lq
parent simulation\rq) as described above, and evolve the system forward
in time using Gadget-2 (see {\it S5)}, following just the growth of
structure in the dark matter for simplicity and speed. Once this
simulation reaches redshift $z=0$, we analyse it and search for a
massive dark matter halo, such as would envelope a massive cluster of
galaxies (or any other structure of interest).

Next we create a multi-mass particle distribution to represent the
unperturbed homogeneous universe, by concentrating lower-mass particles
in the region of interest, with successive layers of more massive
particles away from this region. The original displacements field for
the parent simulation is recreated and the displacements applied
everywhere.  In the region of interest, the inter particle spacing is
smaller, and we can include power on smaller scales than was possible
with the original coarser particle distribution. This small scale power
is generated by creating a periodic cubic Fourier mesh which is placed
around the region of interest. A displacement field is created on this
mesh, with the appropriate power spectrum, taking wave numbers between
the cut-off in power spectrum of the parent simulation and the Nyquist
frequency, $k=\pi/d$, of the region of interest determined by the inter
particle spacing, $d$, there.  Typically the longest waves used in the
small scale power have a maximum wavelength of a tenth of the cubic
Fourier mesh (if the maximum wavelength were to be much larger this
procedure would start becoming inaccurate because periodicity is
enforced by the method over the mesh). The second set of displacements
which represent the additional small scale power, is then applied to
the particles inside the high resolution region only.  The scheme used
to interpolate the displacements to the positions of the particles
takes account of the mass of the particles to avoid aliasing effects
for the most massive particles in the high resolution region.

Having created these initial conditions and run them forward in time, a
new region of interest can be defined for this re-simulation at the
same or different redshift, and a re-simulation of the re-simulation
can be made following the same procedure where the parent simulation is
now identified as the first re-simulation. The simulations described in
this paper apply this procedure four times in order to generate ICs for
the progenitor of a massive cluster of galaxies at redshift $z=0$, and
its surroundings; see ({\it S6}) for further details. The main difference
here is that we used a WDM power spectrum for generating the initial
density field. Note, for simplicity we did not add the thermal
velocities of the WDM particles, hence the free-streaming of the WDM is
mimicked just by their suppression of small-scale power.

The highest resolution achieved in our final simulation is $M_{\rm
  dm}=272.6\,M_\odot$ and $M_{\rm gas}=41.9\,M_\odot$ for dark matter
and gas particles respectively, and the high resolution region has a
Lagrangian radius of about 600 comoving kpc.
\subsection{The simulation code Gadget-2}

Gadget-2 (see {\it S5}) is a Lagrangian code that uses two sets of
particles to represent dark matter and gas. Each gas and dark matter
particle represents a swarm of baryonic particles and gravitinos,
respectively. Because the simulation particles are so much more massive
than the physical particles they represent, we soften the gravitational
forces between simulation particles to minimise numerical
artifacts. Each dark matter particle in the simulation has a position,
velocity, (constant) mass, and gravitational softening length. Gas
particles have in addition a thermal energy and gas smoothing length,
in addition to properties to follow the molecular hydrogen formation
(see next section). Gadget-2 uses the smoothed particle hydrodynamics
(SPH) scheme to evaluate pressure gradients between particles. Once
gravitational and pressure accelerations are computed, the state of
system is marched forward in time. Gadget-2 uses a sophisticated tree
structure to perform these force calculations efficiently on a parallel
computer with message-passing interface (MPI). Full details on Gadget-2
are given in {\it S5}.

\section{Primordial gas chemistry}
In this section, we briefly summarise the basic gas processes that are
important for primordial gas unpolluted by metals; full details are
given in {\it (S6-10)}.

At high redshifts, in the absence of metals, and when the temperature
of gas is lower than $10^4$K (above which atomic hydrogen line cooling
is dominant), the main coolant enabling star formation in the early
dark matter potential wells, is molecular hydrogen, H$_2$. Its
formation has three main channels:

\begin{enumerate} \item the ${\rm H_2^+}$ channel
\begin{eqnarray}
\rm H^+~+~H~   &\rightarrow &\rm ~H_2^+~+~\gamma \nonumber \\
\rm H_2^+~+~H~ &\rightarrow &\rm ~H_2~+\rm~H^+ \nonumber
\end{eqnarray}
\item the H$^-$ channel
\begin{eqnarray}
{\rm H}~+~e^-&\rightarrow&~{\rm H^-~+~\gamma}  \nonumber \\
~~{\rm H^-~+~H}&\rightarrow&~{\rm H_2}~+~e^-   \nonumber
\end{eqnarray}
\item the three-body reaction
\begin{eqnarray}
&{\rm H~+~H~+~H} &\rightarrow{\rm~H_2~+~H},  \nonumber \\
&{\rm H~+~H~+~H_2} &\rightarrow{\rm~H_2~+~H_2}  \nonumber .
\end{eqnarray}
\end{enumerate}
The first channel dominates at high redshift, $z\ge 200$, and the
three-body reactions are important only at high densities (${n_{\rm H}}
\sim 10^8 {\rm cm^{-3}}$); therefore the H$^-$ channel is most relevant
for the initial stages in the formation of the filament discussed in
this paper. Our simulation code follows reactions for $9$ species
$(e^{-}$, ${\rm H}$, ${\rm H^{+}}$, ${\rm He}$, ${\rm He^{+}}$, ${\rm
He^{++}}$, ${\rm H_2}$, ${\rm H_2^+}$, ${\rm H^-}$), and includes all
three H$_2$ formation channels.

\section{Three-dimensional representation of the first filament}
Figure {\it S2} contains projections of the filament shown in Fig.~1
(panels C and D) along all three coordinate directions to illustrate
its three-dimensional cylindrical form. The structure of the filament
on larger scales is illustrated in figure {\it S3}, where we have
transformed the Cartesian axes such that the filament lies along the
$x$-axis. The figure illustrates how straight the filament is on scales
of 100s of parsecs, whereas the tidal field begins to deform the filament
on scales approaching the free-streaming length.


\bigskip
{\large{\bf References and Notes}}

S1. P. J. E. Peebles, {\it Princeton University Press} (1993)

S2. P. Bode, J. P. Ostriker, N. Turok, {\it The Astrophys. J.},{\bf 556},
93 (2001)

S3. Y. B. Zel'Dovich, {\it Astron. \& Astrophys.},{\bf 5}, 84 (1970)

S4. G. Efstathiou, M. Davis, S. D. M. White, C. S. Frenk, {\it The
  Astrophys. J.}, {\bf 57}, 241 (1985)

S5. V. Springel, {\it Mon. Not. R. Astron. Soc.},  {\bf 364}, 1105
(2005)

S6. L. Gao, S. D. M. White, A. Jenkins, C. S. Frenk, V. Springel, {\it
  Mon. Not. R. Astron. Soc.}, {\bf 363}, 379 (2005)

S7. F. Palla, E. E. Salpeter, S. W. Stahler,  {\it The Astrophys. J.},
{\bf 271}, 632 (1985)

S8. D. Galli, F. Palla, {\it Astron. \& Astrophys.}, {\bf 335}, 403
(1998)

S9. N. Yoshida, K. Omukai, L. Hernguist,T. Abel, {\it The Astrophys. J.},
{\bf 652}, 6 (2006)

S10. L. Gao, N. Yoshida, T. Abel, C. S. Frenk, A. Jenkins, V. Springel,
{\it Mon. Not. R. Astron. Soc.}, {\bf 378}, 449 (2007)

\begin{figure*}
\hspace{2cm}\resizebox{10cm}{!}{\includegraphics{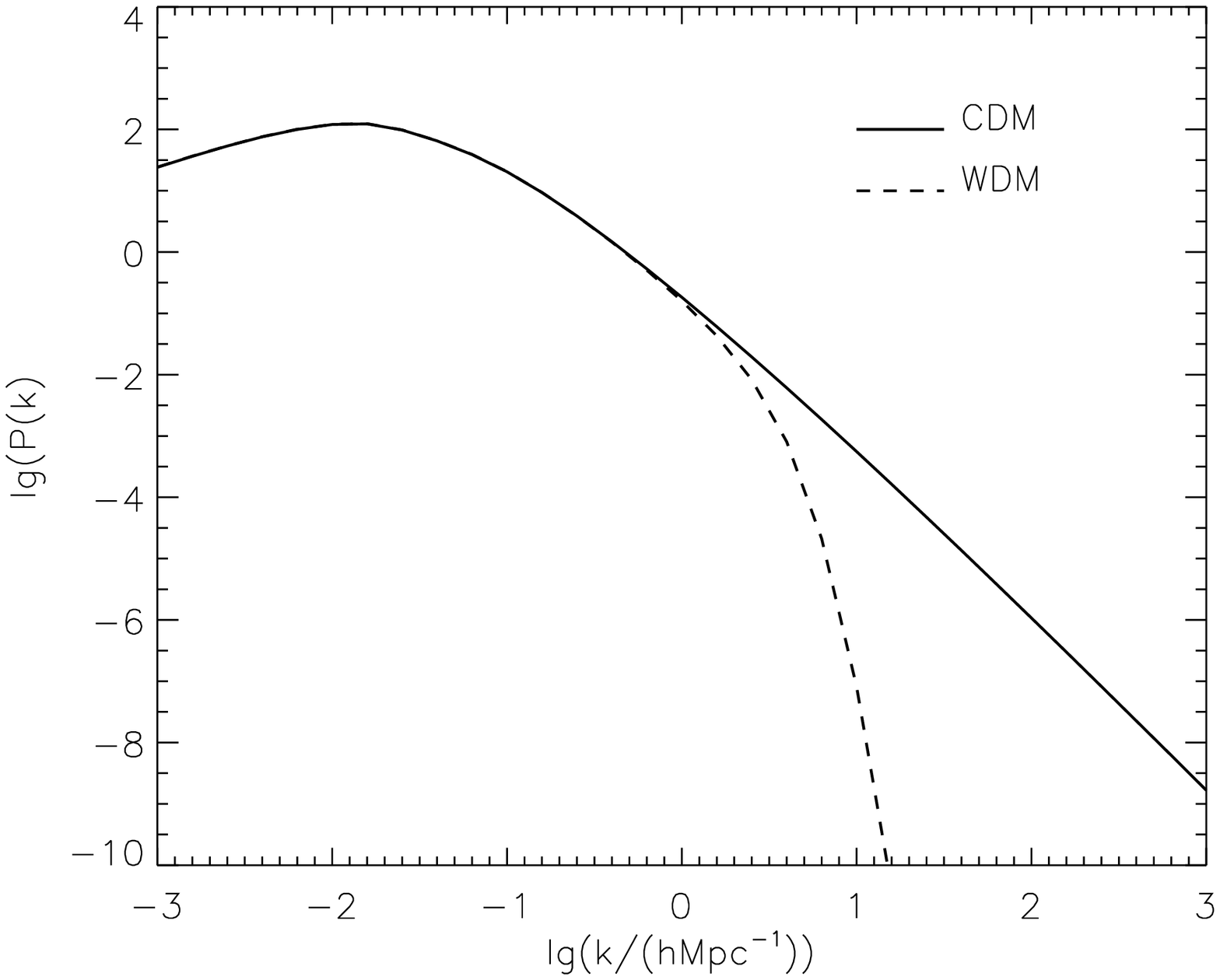}}
\\
\noindent Figure S1:
Power-spectrum for CDM (drawn line) and WDM (dashed
line). Free-streaming of the warm dark matter particle (a gravitino
with mass $m_{\rm WDM}=3$~keV) exponentially suppresses power below a
scale $\sim 100$~kpc.
\end{figure*}

\begin{figure*}
\vspace{-2cm}
\resizebox{8cm}{!}{\includegraphics{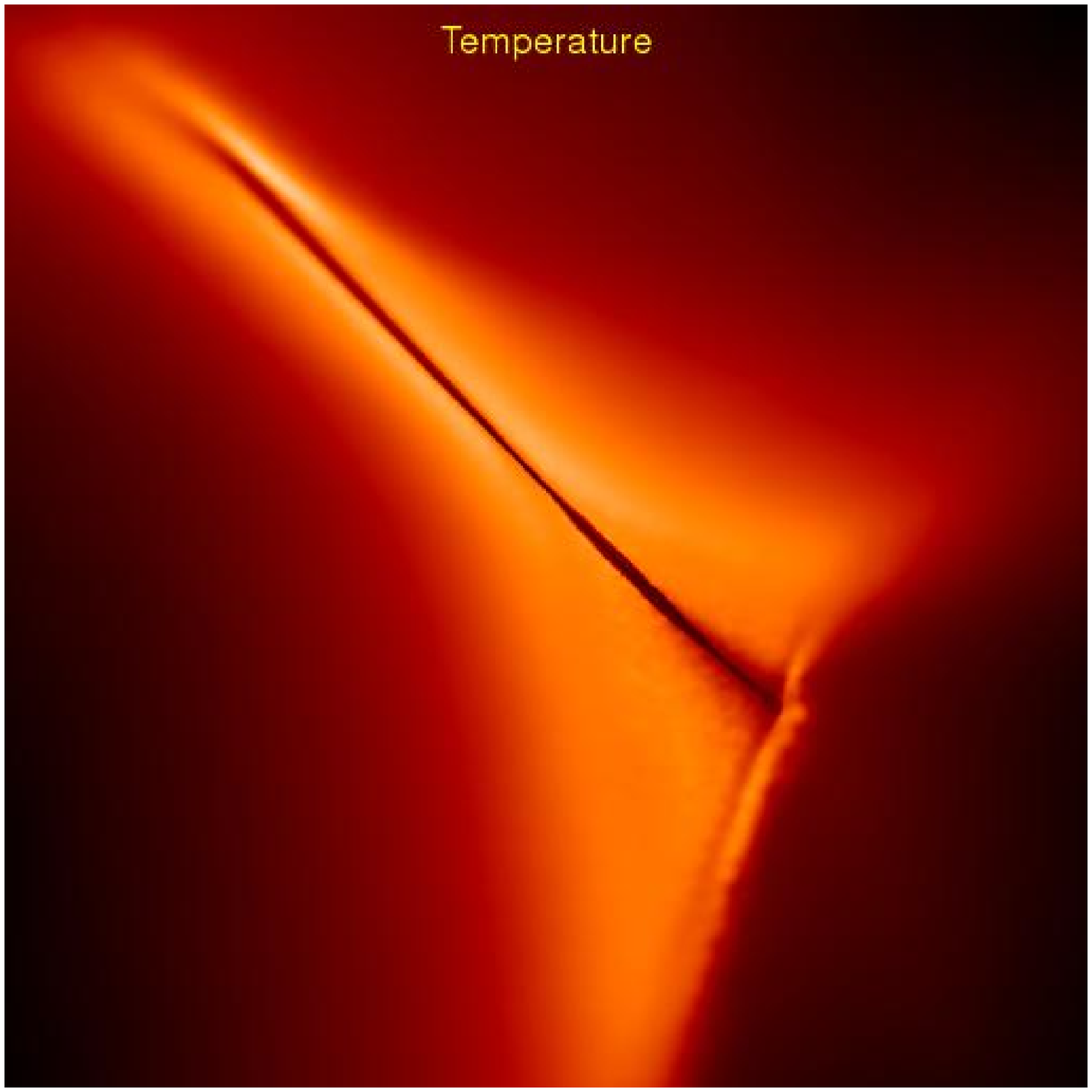}}
\hspace{0.13cm}\resizebox{8cm}{!}{\includegraphics{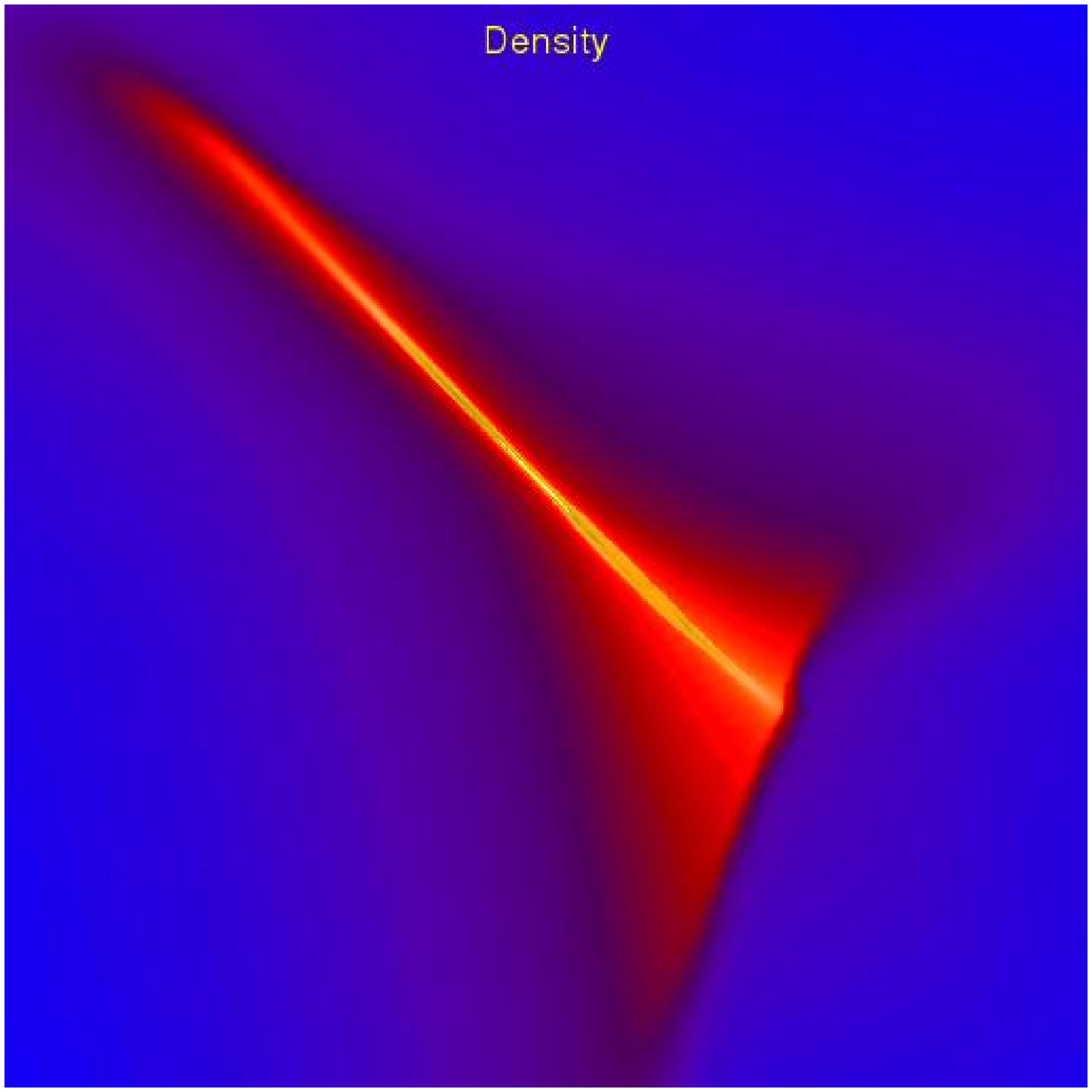}}
\hspace{0.13cm}\resizebox{8cm}{!}{\includegraphics{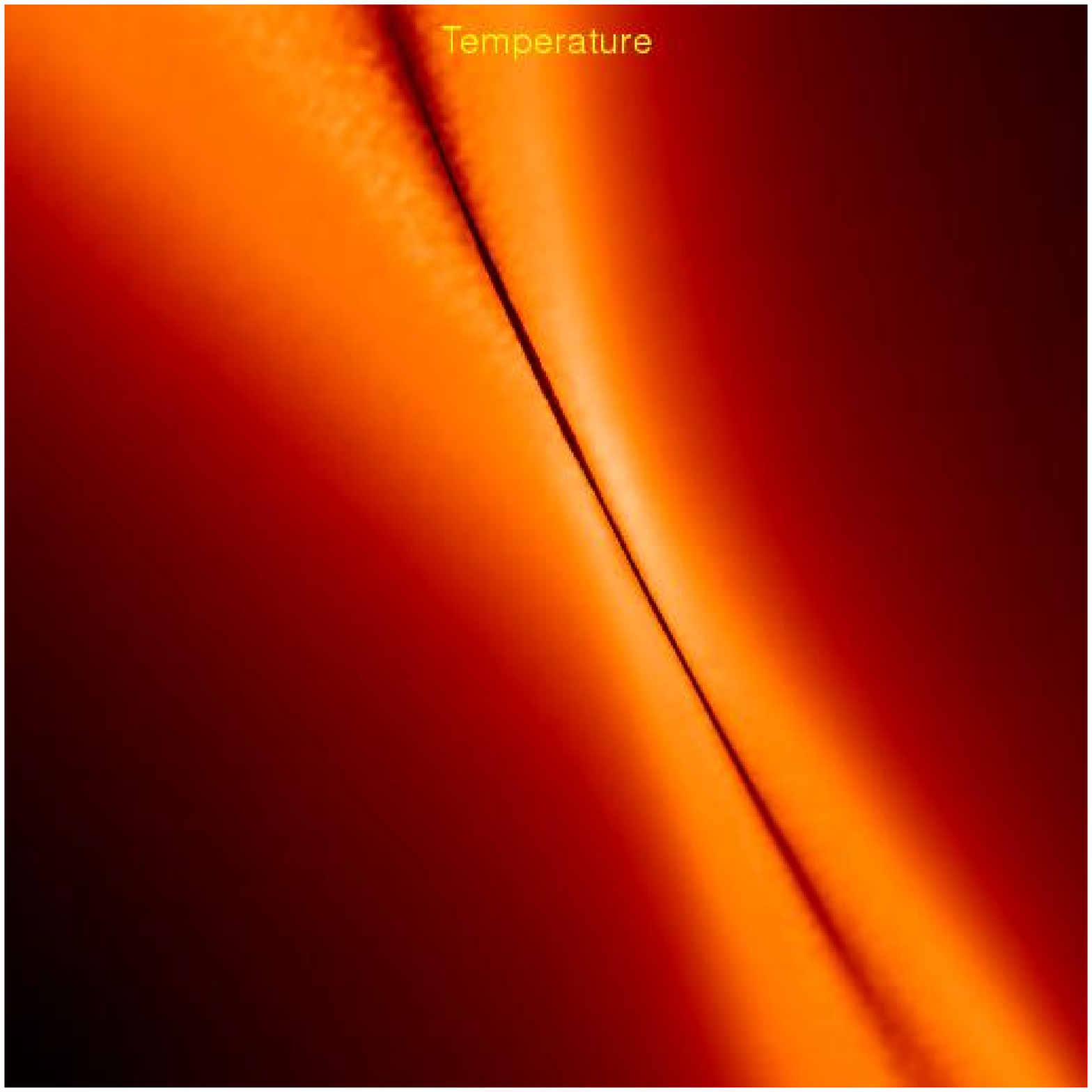}}
\hspace{0.13cm}\resizebox{8cm}{!}{\includegraphics{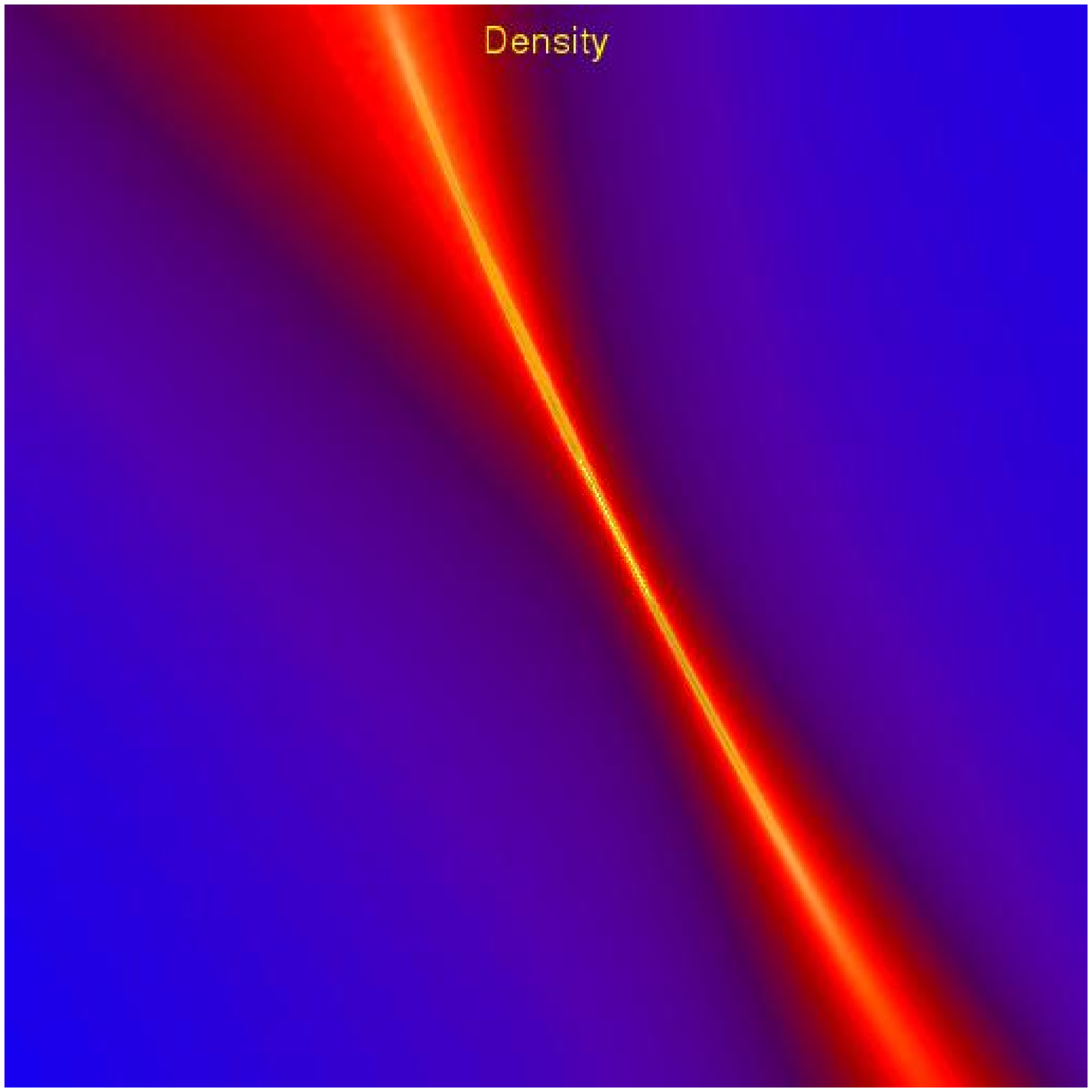}}
\hspace{0.13cm}\resizebox{8cm}{!}{\includegraphics{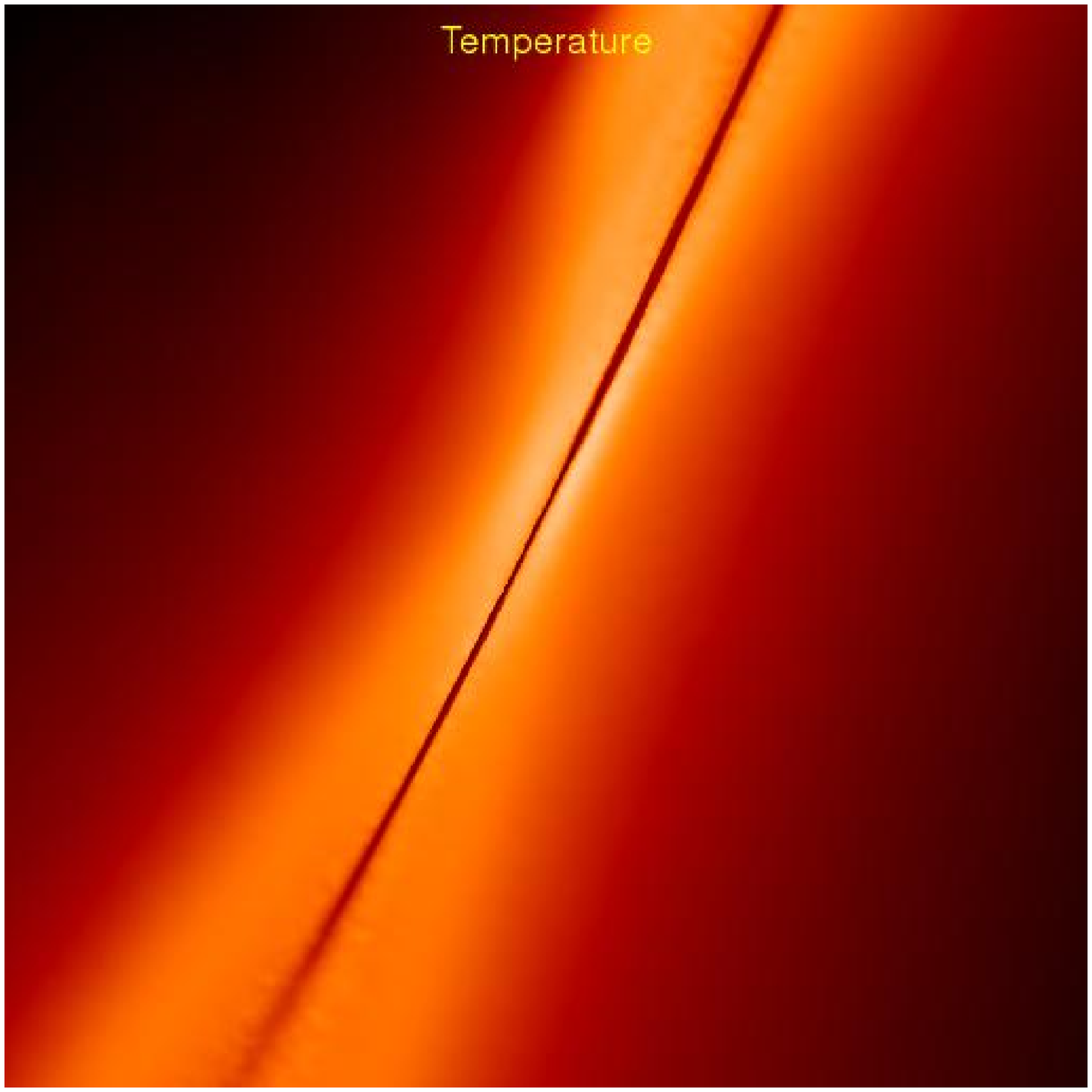}}
\hspace{0.13cm}\resizebox{8cm}{!}{\includegraphics{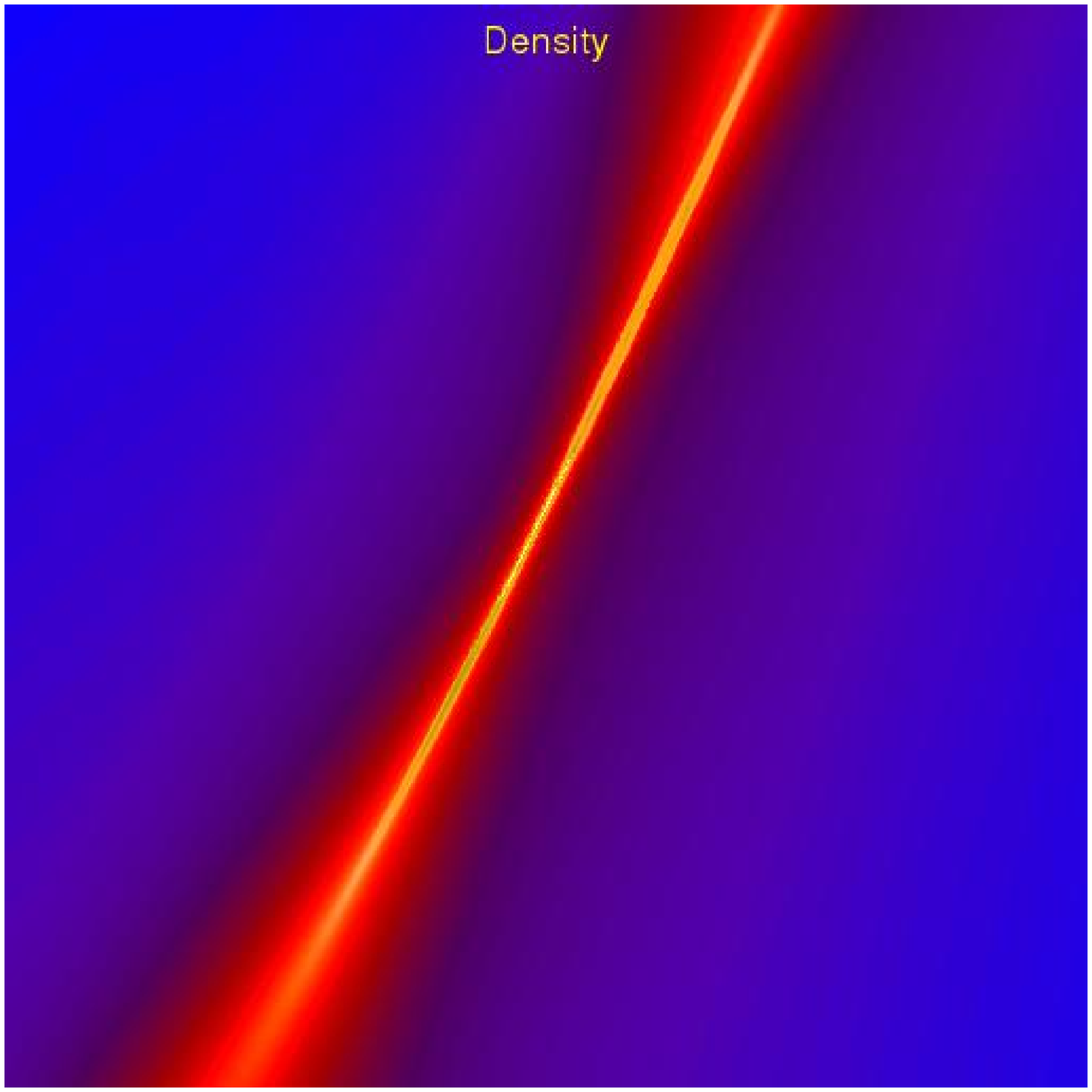}}
\\
\noindent Figure S2:

Gas temperature (left hand panels) and density (right hand panels) of
the filament shown in Fig.~1 (panels C and D) along three orthogonal
projections. Each panel is 800 (physical) parsec on a side.

\end{figure*}

\begin{figure*}
\vspace{-2cm}
\resizebox{8cm}{!}{\includegraphics{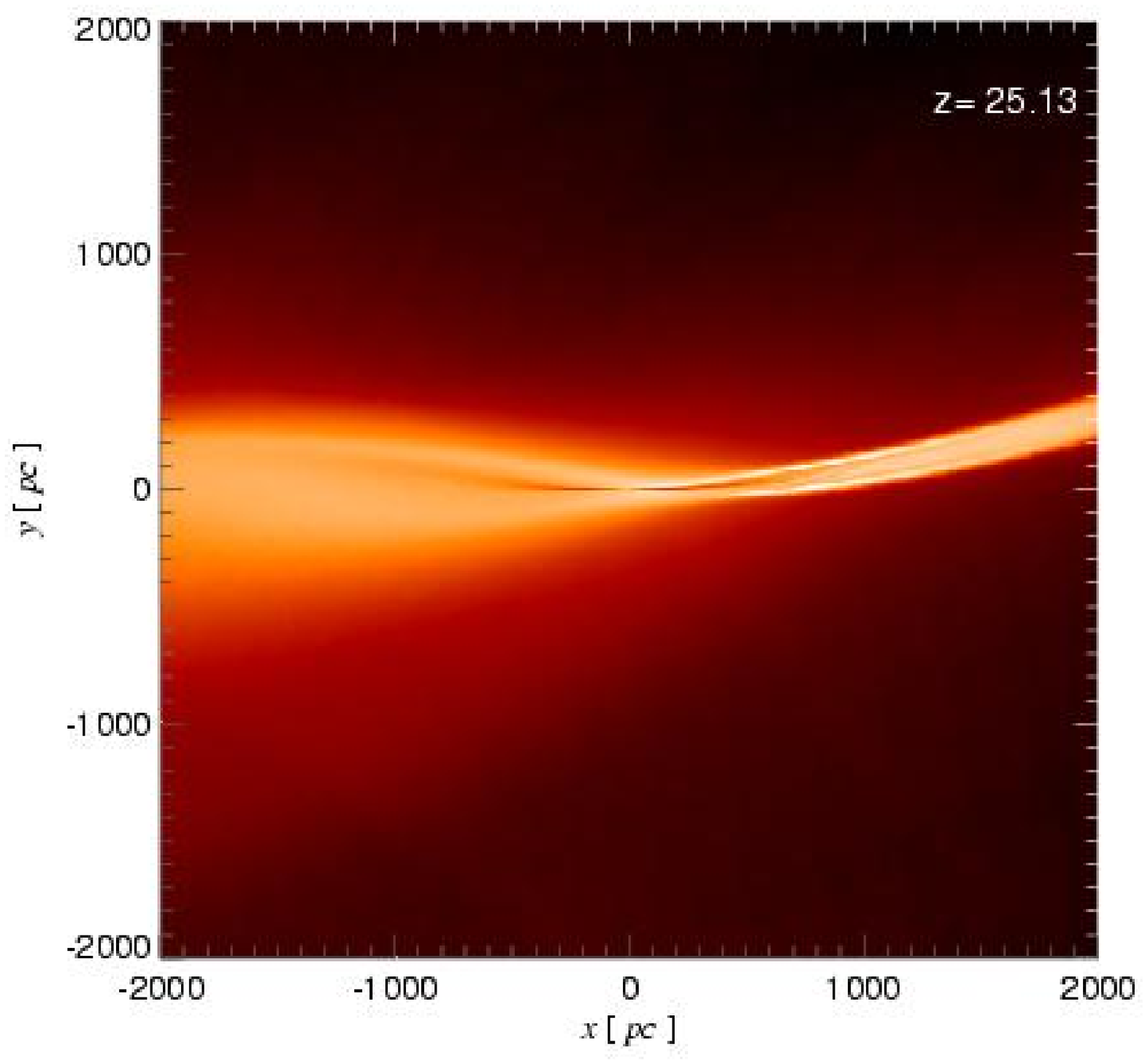}}
\resizebox{8cm}{!}{\includegraphics{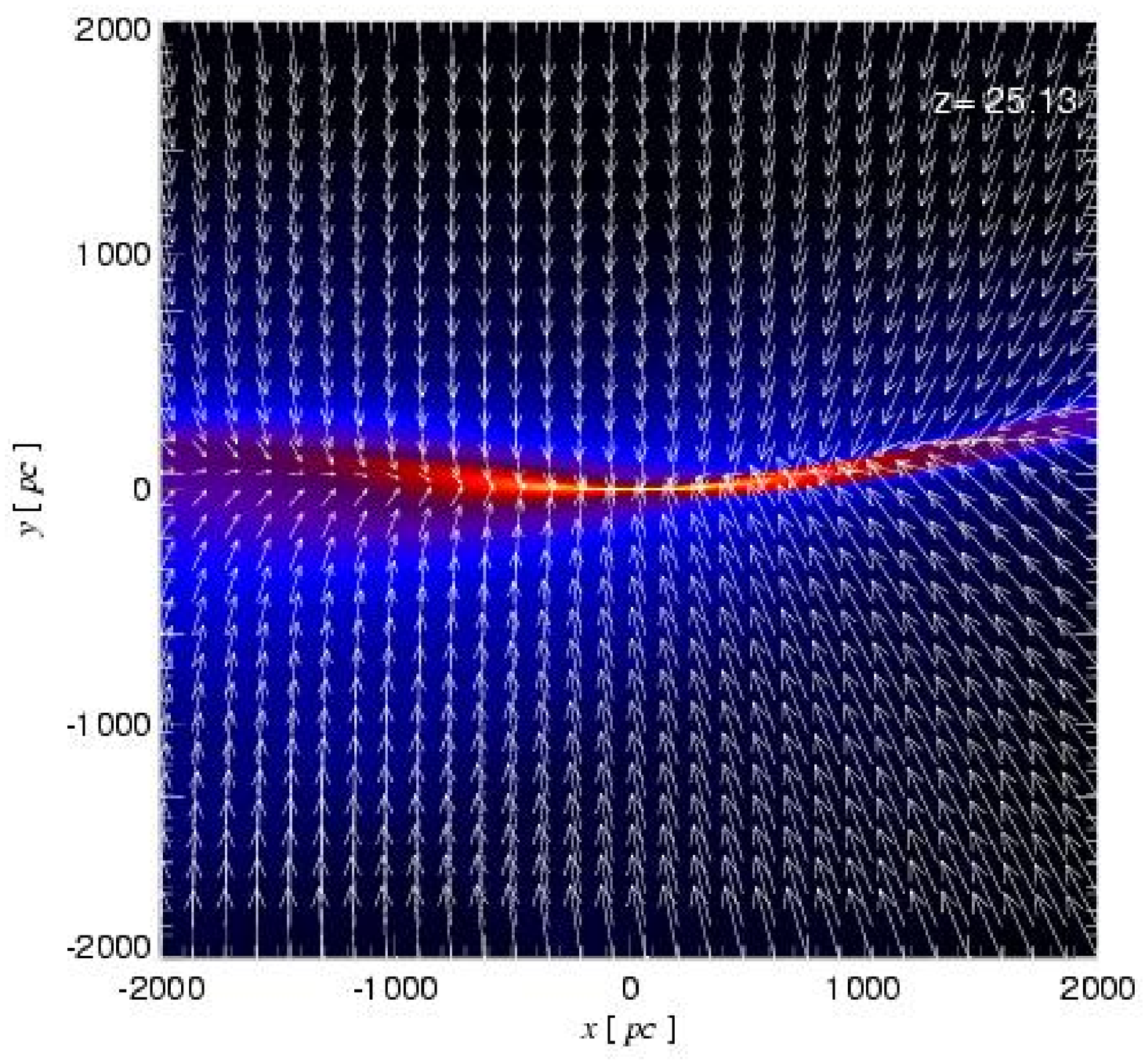}}
\resizebox{8cm}{!}{\includegraphics{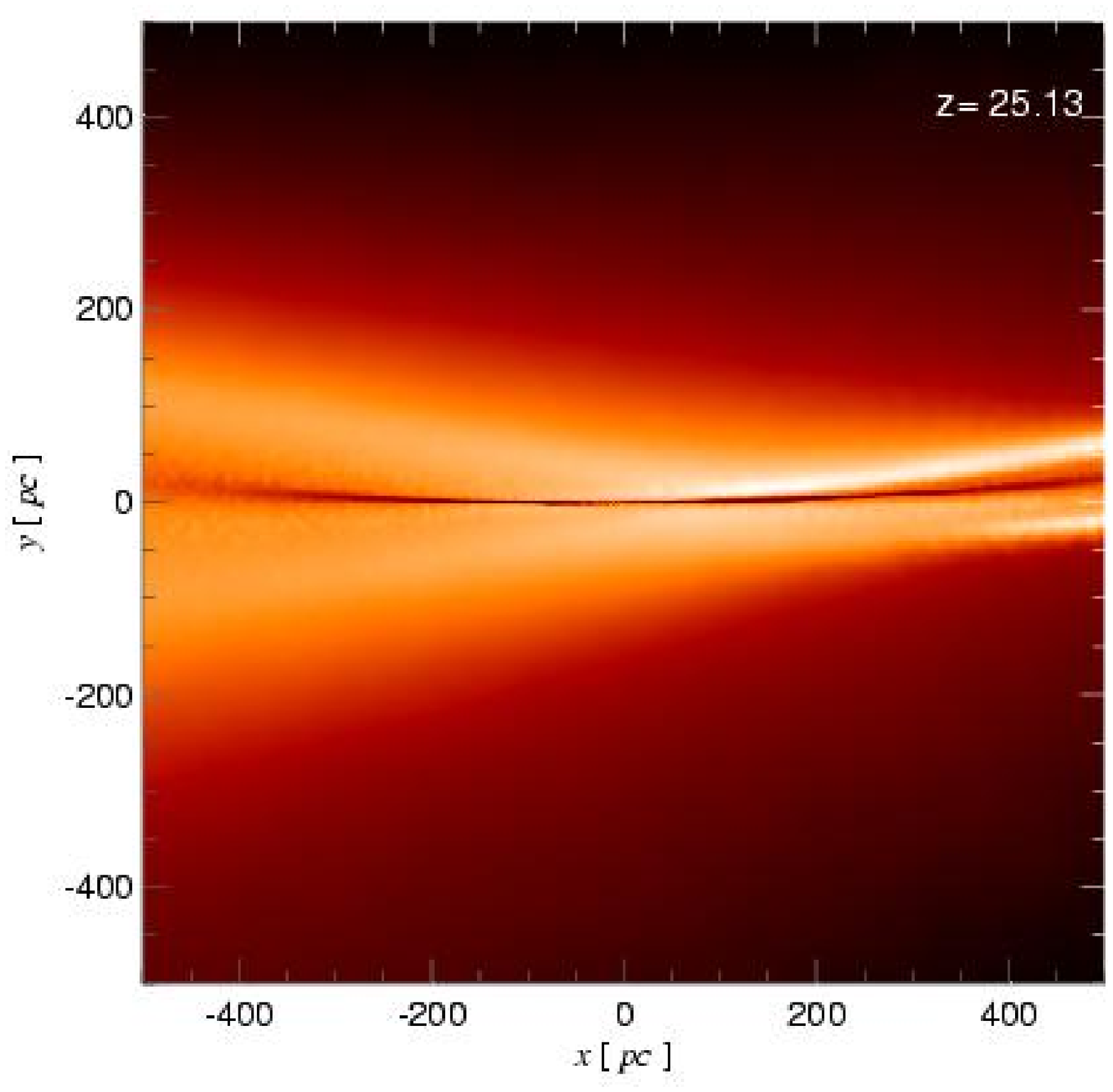}}
\resizebox{8cm}{!}{\includegraphics{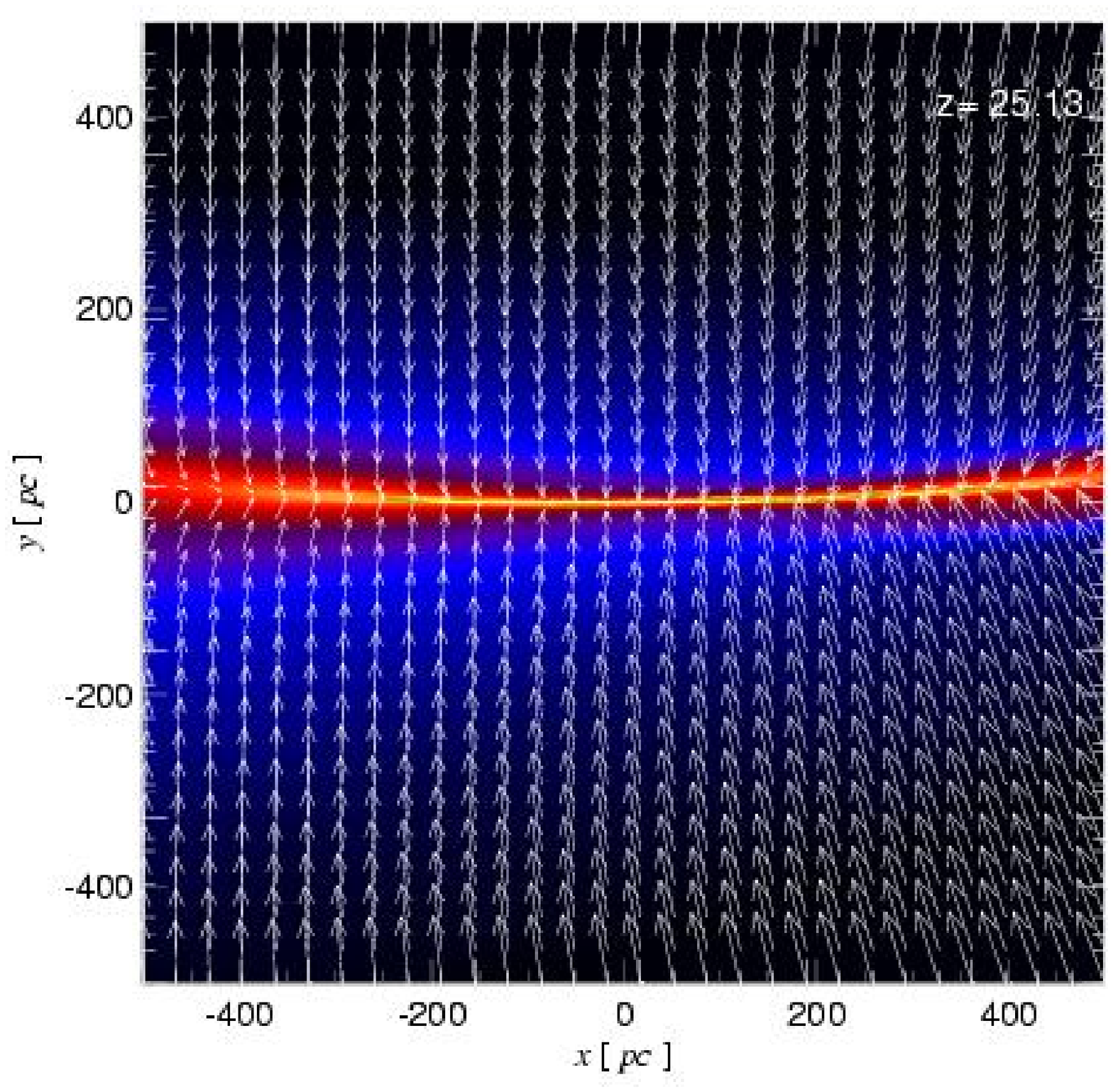}}
\resizebox{8cm}{!}{\includegraphics{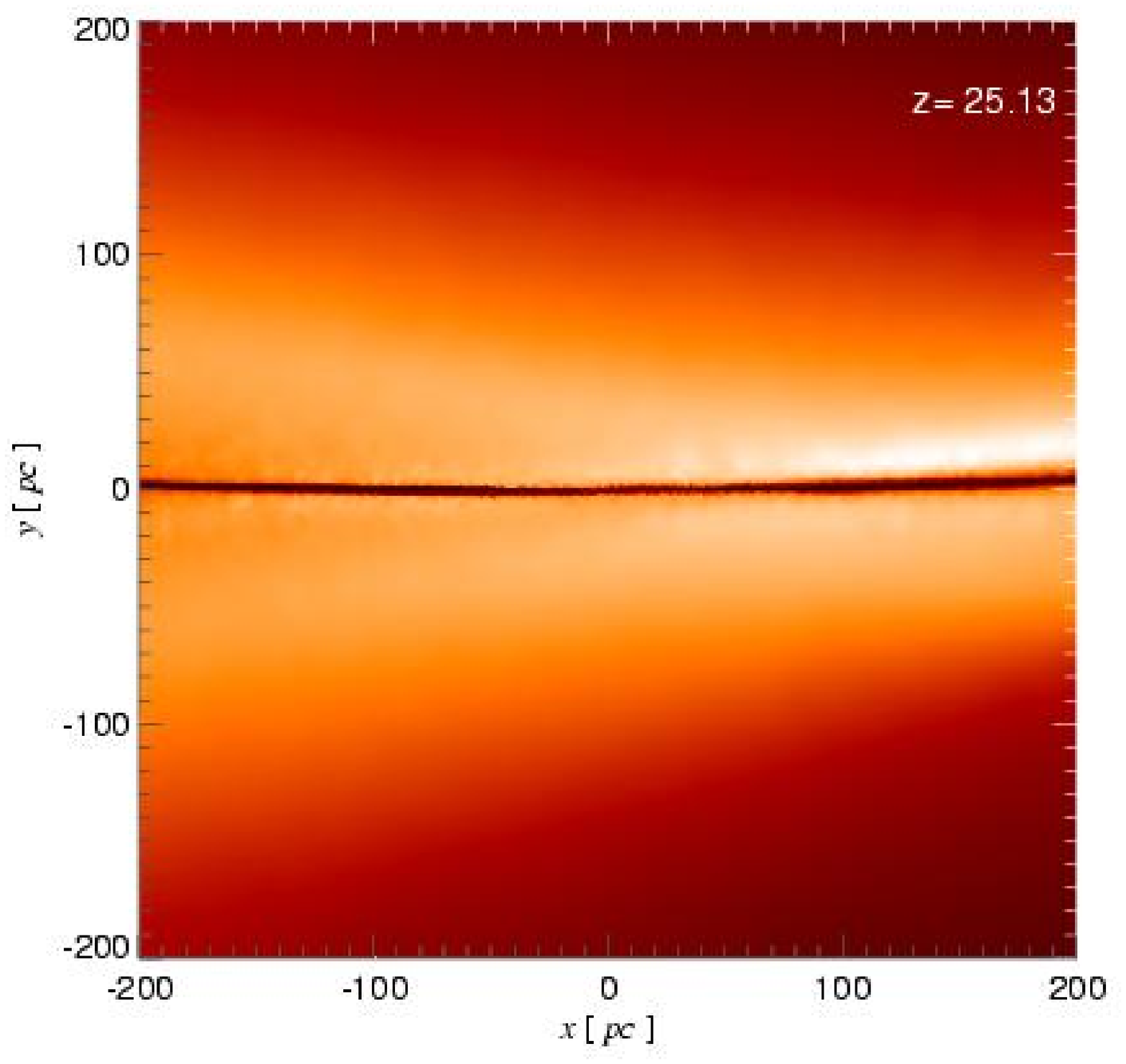}}
\resizebox{8cm}{!}{\includegraphics{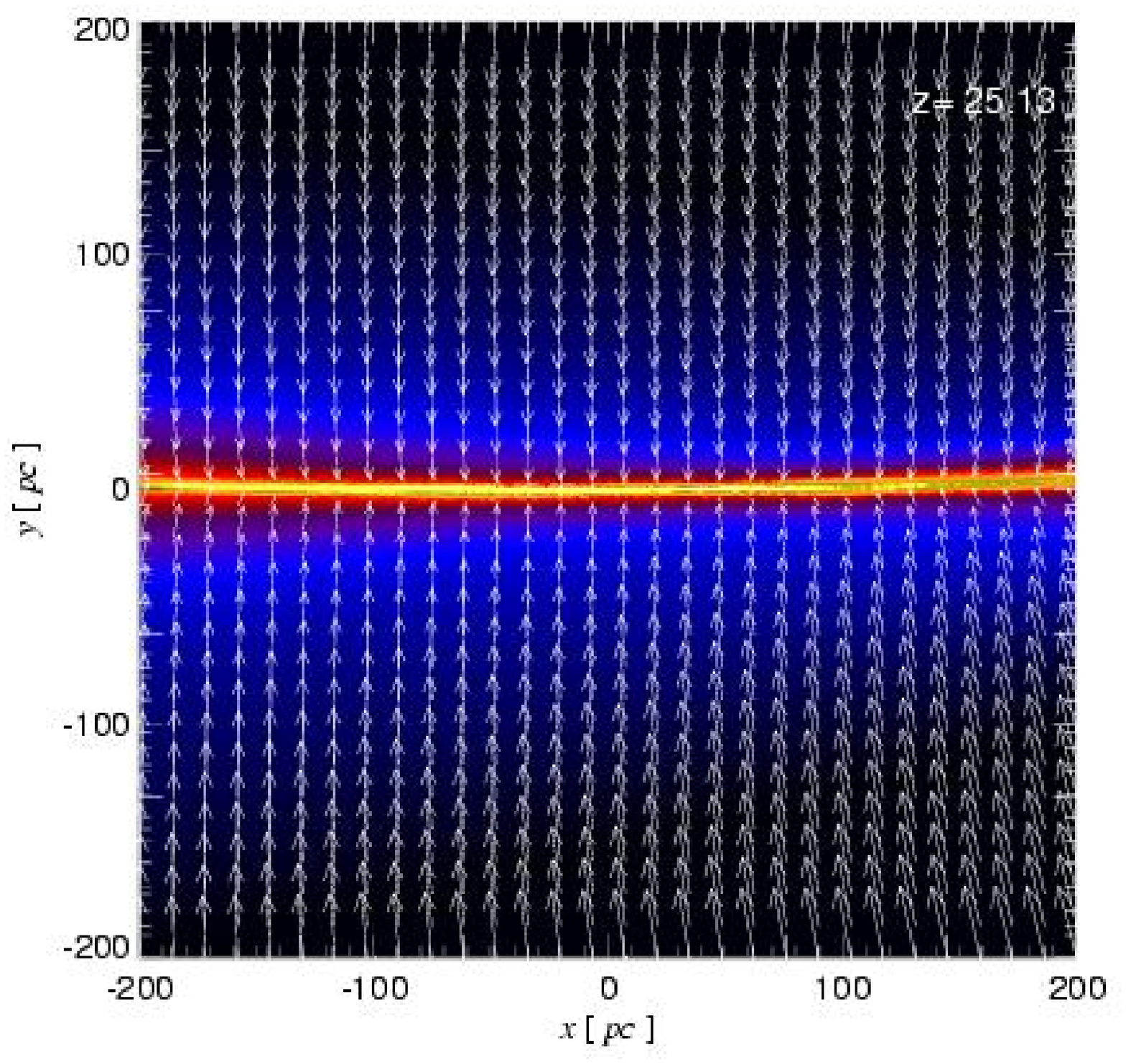}}
\\
\noindent Figure S3:

Gas temperature (left hand panels) and density (right hand panels) of
the filament shown in Fig.~1 (panels C and D), after rotating the
filament so that its long axis lies along the $x$-axis. Panels from
top to bottom zoom out from 4~kpc to 400~pc. The accretion velocities
of the gas onto the filament are shown by the velocity vectors.

\end{figure*}


\begin{thebibliography}{10}

\bibitem{M99}
B. Moore, S. Ghigna, F. Governato, G. Lake, T. Quinn T., J. Stadel,
P. Tozzi, {\it The Astrophys. J.}, {\bf 524}, L19 (1999)

\bibitem{DH01}
J. J. Dalcanton, C. J. Hogan, {\it The Astrophys. J.}, {\bf 561}, 35
(2001)

\bibitem{Bode01}
P. Bode, J. P. Ostriker, N. Turok, {\it The Astrophys. J.},{\bf 556},
93 (2001)

\bibitem{ABN02}
T. Abel, G. L. Bryan, M. L. Norman,{\it Science}, {\bf 295}, 93 (2002)

\bibitem{BCL02}
V. Bromm, P. S. Coppi \& R. B. Larson, {\it The Astrophys. J.}, {\bf
  564}, 23 (2002)

\bibitem{Yoshida06}
N. Yoshida, K. Omukai, L. Hernguist,T. Abel, {\it The Astrophys. J.},
{\bf 652}, 6 (2006)


\bibitem{G07}
L. Gao, N. Yoshida, T. Abel, C. S. Frenk, A. Jenkins, V. Springel,
{\it Mon. Not. R. Astron. Soc.}, {\bf 378}, 449 (2007)

\bibitem{G05}
L. Gao , S. D. M. White, A. Jenkins, C. S. Frenk, V. Springel, {\it
  Mon. Not. R. Astron. Soc.}, {\bf 363}, 379 (2005)

\bibitem{SO} See supporting material on Science on line.

\bibitem{S05}
V. Springel, {\it Mon. Not. R. Astron. Soc.},  {\bf 364}, 1105 (2005)

\bibitem{GP98}
D. Galli, F. Palla, {\it Astron. \& Astrophys.}, {\bf 335}, 403 (1998)

\bibitem{bhs05}
G. Bertone, D. Hooper, J. Silk, Particle dark matter: evidence,
candidates and constraints, {\it Physics Report}, {\bf 405}, 279 (2005)

\bibitem{neutrinos}
S. Dodelson, L. M. Widrow, {\it Phys. Rev. Lett.},  {\bf 72}, 17 (1994)

\bibitem{Viel06}
M. Viel, J. Lesgourgues, M. G. Haehnelt, S. Matarrese, A. Riotto, {\it
  Phys. Rev. Lett.}, {\bf 97}, 071301 (2006)

\bibitem{Seljak06}
U. Seljak., A. Makarov, P. McDonald, H. Trac, {\it
  Phys. Rev. Lett.},{\bf 97}, 191303 (2006)


\bibitem{wmap3}
D. N. Spergel {\it et al.}, {\it The Astrophys. J.} in press, preprint
astro-ph/0603449 (2007)

\bibitem{Goetz02}
M. G\"otz, J. Sommer-Larsen, {\it Astron. \& S. Sci.}, {\bf 281}, 415
(2002)

\bibitem{W07}
J. Wang, S. D. M. White, preprint, astro-ph/0702575 (2007)

\bibitem{HD}
J. L. Johnson, V. Bromm, {\it Mon. Not. R. Astron. Soc.}, {\bf
  366}, 247 (2006)

\bibitem{Yoshida07}
N. Yoshida, S.~P. Oh, T. Kitayama, L. Hernquist, {\it The
  Astrophys. J.}, {\bf 663}, 687 (2007)

\bibitem{L85}
R. B. Larson, {\it Mon. Not. R. Astron. Soc.}, {\bf 214}, 379 (1985)

\bibitem{IM97}
S. Inutsuka, S. M. Miyama, {\it The Astrophys. J.}, {\bf 480},
681 (1997)

\bibitem{NU02}
F. Nakamura, M. Umemura, {\it The Astrophys. J.}, {\bf 569}, 549 (2002)

\bibitem{Omukai98} 
K. Omukai, R. Nishi, {\it The Astrophys. J.}, {\bf 508}, 141  (1998)

\bibitem{Larson05}
R.~B. Larson, {\it Mon. Not. R. Astron. Soc.}, {\bf 359}, 211 (2005)

\bibitem{CH02}
N. Christlieb, M. S. Bessell, T. C. Beers, B. Gustafsson, A. Korn,
P. S. Barklem, T. Karlsson, M. Mizuno-Wiedner, S. Rossi, {\it
  Nature}, {\bf 419}, 904 (2002)

\bibitem{F05}
Frebel {\it et al.}, {\it Nature}, {\bf 434}, 871 (2005)


\bibitem{BC05}
T. C. Beers, N. Christlieb, {\it Annu. Rev. Astron. Astrophys.}, {\bf
  43}, 531 (2005)


\bibitem{Y03}
N. Yoshida, A. Sokasian, L. Hernquist, V. Springe, {\it The
  Astrophys. J.}, {\bf 591}, 1 (2003)

\bibitem{ON06}
B. W. O'Shea, M. L. Norman, {\it The Astrophys. J.}, {\bf 648}, 31 (2006)
\end{thebibliography}
\end{document}